\documentclass[RNAAS]{aastex63}

\usepackage{hyperref}
\usepackage{color}

\newcommand{\chandra}{{\sl Chandra}}
\newcommand{\cstat}{{\tt cstat}}
\graphicspath{{./}{figures/}}
\begin{document}

\title{Spectral fit residuals as an indicator to increase model complexity}

\correspondingauthor{Anshuman Acharya}
\email{anshuman@mpa-garching.mpg.de}

\author[0000-0003-3401-4884]{Anshuman Acharya}
\affiliation{Max Planck Institut für Astrophysik, \\ 85748 - Garching, Germany.}

\author[0000-0002-3869-7996]{Vinay L.\ Kashyap}
\affiliation{Center for Astrophysics $|$ Harvard \& Smithsonian \\60 Garden St.\ MS-70, Cambridge MA 02138, USA}


\begin{abstract}
Spectral fitting of X-ray data usually involves minimizing statistics like the chi-square and the Cash statistic. Here we discuss their limitations and introduce two measures based on the cumulative sum (CuSum) of model residuals to evaluate whether model complexity could be increased: the percentage of bins exceeding a nominal threshold in a CuSum array (pct$_{CuSum}$), and the excess area under the CuSum compared to the nominal (p$_\textit{area}$).  We demonstrate their use with an application to a {\sl Chandra} ACIS spectral fit.
\end{abstract}

\section{Introduction}
Spectral fitting of X-ray data has been usually done by minimizing a statistic like $\chi^2$ or {\tt cstat} \citep{1979ApJ...228..939C,1976ApJ...210..642A}.  The value of the statistic is useful to evaluate the goodness of the fit, which is also used as a stopping rule in evaluating model complexity.  Fits are deemed acceptable if $\chi^2/\nu < 1 + 3\cdot\sqrt{\frac{2}{\nu}}$ (where $\nu$ is the degrees of freedom in the fit) or $\Delta${\tt cstat}$<3\cdot\sigma_{\tt cstat}$ (where $\sigma^2_{\tt cstat}$ is the estimated variance for the nominal model as given by \citealt{Kaastra_2017}).  When the fitting statistics or F-test values drop below the threshold defined by the null distributions, it is commonly acknowledged that further increases in model complexity (or the number of free parameters in the fit) are statistically unsupported.  However, these measures are {\sl global} measures of fit, and leave contiguous ranges in data space with correlated deviations in the residuals.  Here we introduce summary statistics that identify the presence of such structures in the residuals and describe how to use them to decide whether the model complexity should be increased.

We use a \chandra/ACIS-S spectrum of the corona of an exoplanet hosting star \citep[HD\,179949, ObsID 6122;][]{Acharya_2023} to illustrate the measures (see Figure~\ref{fig:1}). The data and the spectral fits made with different emission models are shown in the left column of the Figure.  The models increase in complexity from top to bottom (see \citealt{Acharya_2023} for details), with none of the statistical measures ({\tt cstat}, pct$_{CuSum}$, p$_\textit{area}$) acceptable for the simplest model, {\tt cstat} being acceptable for the model in the middle row, and all measures acceptable for the model in the bottom row.  Below, we define pct$_{CuSum}$ and p$_\textit{area}$, describe their rationale, and how to compute them and their null distributions.

\section{CuSum: Definition, Calibration, and Significance}

Spectral fit residuals can be expressed as
    $$\{r_i(\hat{\theta}) = (m_i(\hat{\theta})-c_i), ~~~ i=1,\ldots,N\} \,,$$
with $c_i$ the observed counts in bin $i$ of $N$ and $m_i$ the predicted model counts  in the same bin, generated for an astrophysical model defined by the best-fit parameters $\hat{\theta}$.  The cumulative sum (CuSum) at the $j^{\rm th}$ bin is
\begin{equation}
    {\rm CuSum}(\hat{\theta})_j = \sum_{i=1}^j r_i(\hat{\theta}) \,.
\end{equation}

While the ordering of indices can be reversed, CuSum preserves the sequential order of the bins, thus incorporating extra information ignored in $\chi^2$ or \cstat\ statistics.
The null distribution needed to evaluate the quality of the CuSum of a best-fit can be built in one of two ways:
\hfil\break
$-$ generate $K$ mock datasets using the {\tt fake\_pha()} function in Sherpa, and fit using the same model, such that for each mock data set $k$, we obtain corresponding best-fit parameters $\theta^k$; or
\hfil\break
$-$ obtain $K$ post burn-in draws $\{\theta^k\,,~k=1,\ldots,K\}$ as iterations during a Markov Chain Monte Carlo (MCMC) fit using the {\tt get\_draws()} tool in Sherpa \citep[based on {\tt BLoCXS};][]{van_Dyk_2001}.
\hfil\break
Each $\theta^k$ yields a CuSum array $\{{\rm CuSum}(\theta^{k})_j\,,~j=1,\ldots,N\}$, and the resulting sample 
\begin{equation}
    \{{\rm CuSum}(\theta^k)_j\,,~k=1,\ldots,K,~j=1,\ldots,N\}
    \label{eq:nulldist}
\end{equation} defines the null distribution.

This null distribution characterizes the strength of the deviations present in ${\rm CuSum}(\hat{\theta})_j$.  We construct two statistics that probe deviation at different scales, pct$_{CuSum}$ to detect biases in the model continuum and p$_\textit{area}$ to detect presence/absence of narrow lines.

\begin{enumerate}
    \item pct$_{CuSum}$: We compute the equal-tail 90\% point-wise range $[C^{05}_J,C^{95}_J]$ of ${\rm CuSum}(\theta^k)_{j=J}$ for each detector channel $J$ in the specified passband.
    Then we evaluate the percentage of bins, pct$_{CuSum}$, where the observed CuSum exceeds the 90\% bounds, i.e., if $n$ channels have ${\rm CuSum}(\hat{\theta})_J < C^{05}_J$ or ${\rm CuSum}(\hat{\theta})_J > C^{95}_J$, then pct$_{CuSum}=100\cdot\frac{n}{N}$.  If pct$_{CuSum}{\gtrsim}10$\%, the model used is considered inadequate and we suggest that more complex models with more free parameters should be explored.  If, on the other hand, pct$_{CuSum}{\ll}10$\% this is a sign of overfitting, and thus less complex models are favored (see the middle column of Figure~\ref{fig:1}).  
    \item p$_\textit{area}$: For each of the $k=1,\ldots,K$ CuSum arrays in the null, we calculate the total area that falls outside the 5-95\% bounds across all the channels,
    \begin{equation}
        {\rm area}^k = \sum_J Z^+_J ({\rm CuSum}({\theta^k})_J - C^{95}_J) + Z^-_J (C^{05}_J - {\rm CuSum}({\theta^k})_J) \,,
        \label{eq:parea_sim}
    \end{equation}
    where the indicator indices $Z^+ = Z^- = 1$ if the corresponding terms in the summation are $+$ve, and $0$ otherwise.  We then compute a similar area measure for the best-fit spectrum,
    \begin{equation}
        {\rm area}(\hat{\theta}) = \sum_J Z^+_J ({\rm CuSum}(\hat{\theta})_J - C^{95}_J) + Z^-_J (C^{05}_J - {\rm CuSum}(\hat{\theta})_J) \,.
        \label{eq:parea_obs}
    \end{equation}
    The set $\{{\rm area}^k\,,~k=1,\ldots,K\}$ now forms a null distribution of excess areas to which the value of ${\rm area}(\hat{\theta})$ can be compared to obtain a $p$-value.  The statistic p$_\textit{area}$ is the probability that ${\rm area}(\hat{\theta})>{\rm area}^k$.  We consider the best-fit model an inadequate fit if p$_{area}{\ll}0.05$, as an indication that a large deviation is present in ${\rm CuSum}(\hat{\theta})_j$ (see the right column of Figure~\ref{fig:1}).
    This process is similar to the posterior predictive $p$-value calibration procedure described by \citet{2002ApJ...571..545P}.
\end{enumerate}

\section{Application}

\begin{figure*}[!htbp]
\centering
\begin{tabular}{lll}
\includegraphics[width=0.3\textwidth,height=\textheight,keepaspectratio]{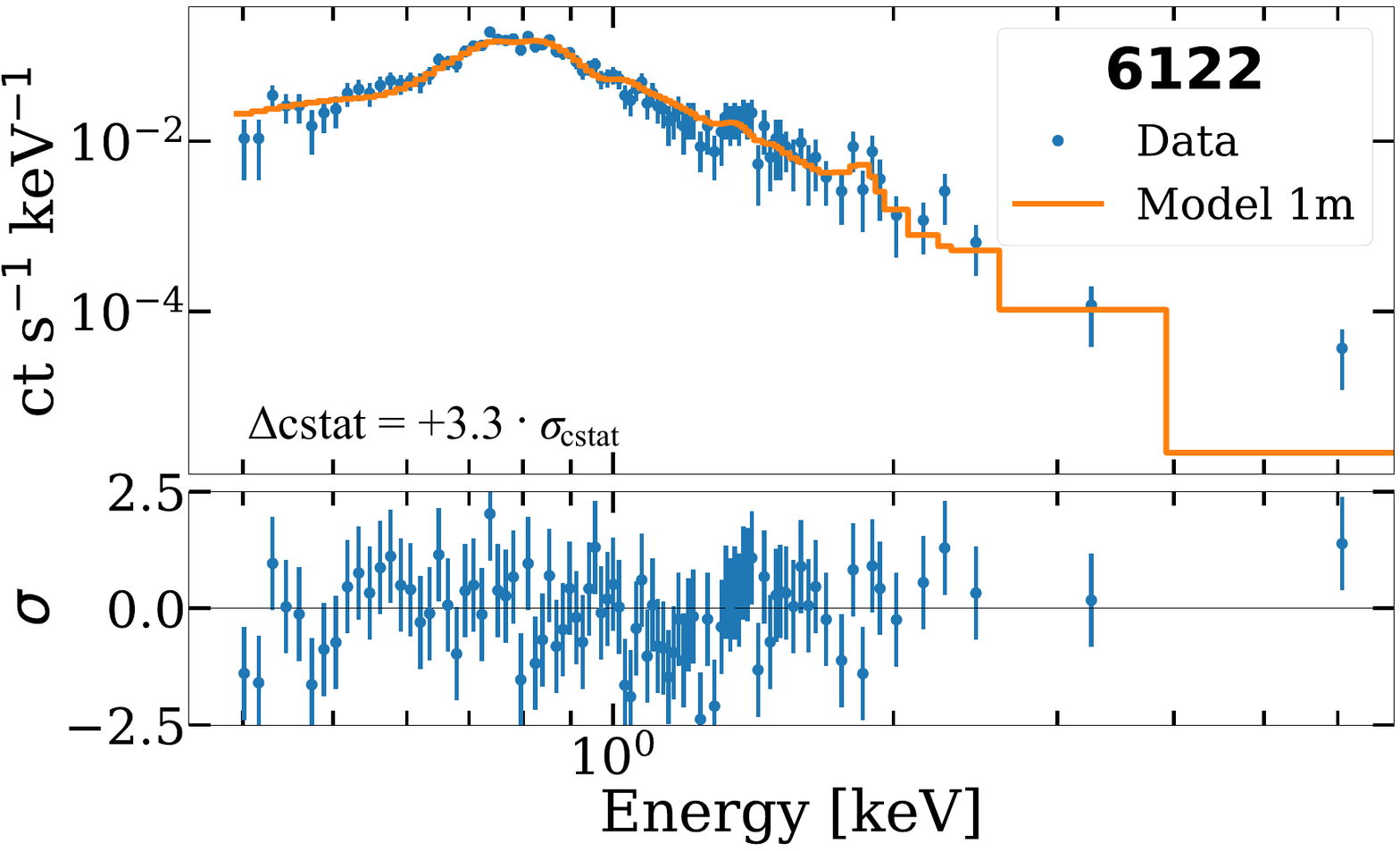} &
\includegraphics[width=0.33\textwidth,height=\textheight,keepaspectratio]{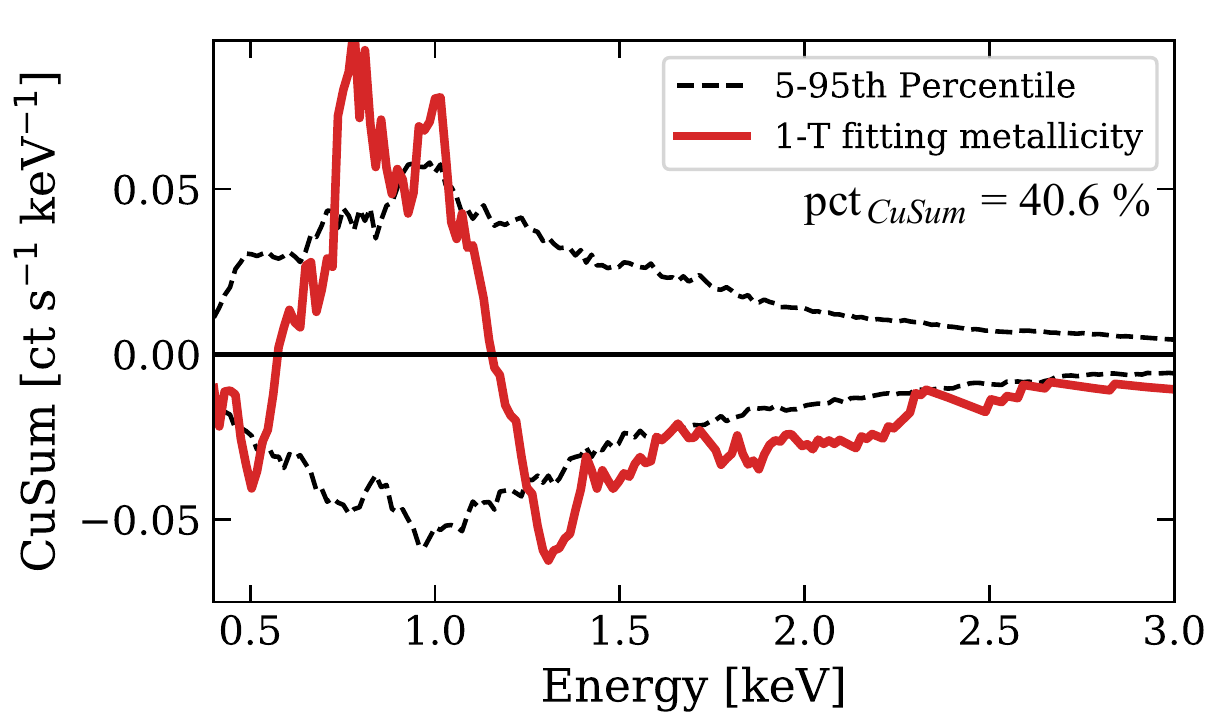} &
\includegraphics[width=0.3\textwidth,height=\textheight,keepaspectratio]{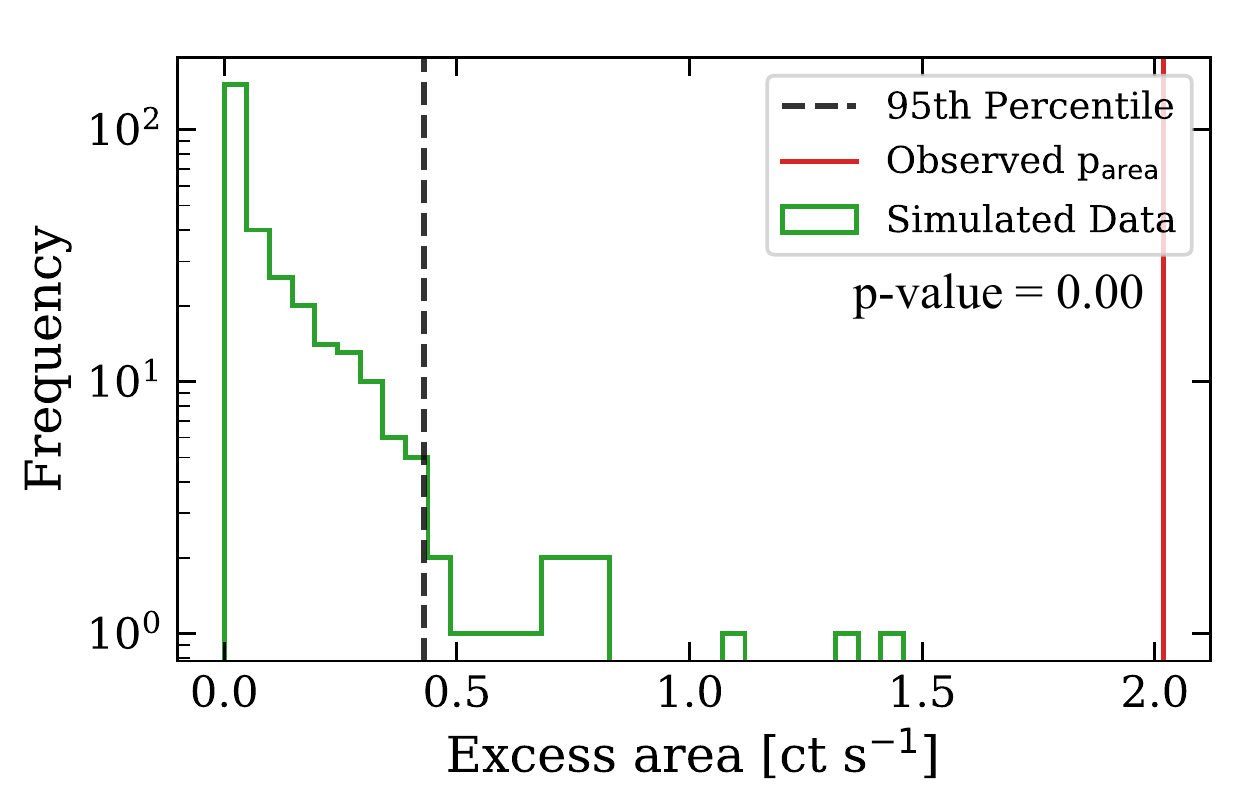} \\
\includegraphics[width=0.3\textwidth,height=\textheight,keepaspectratio]{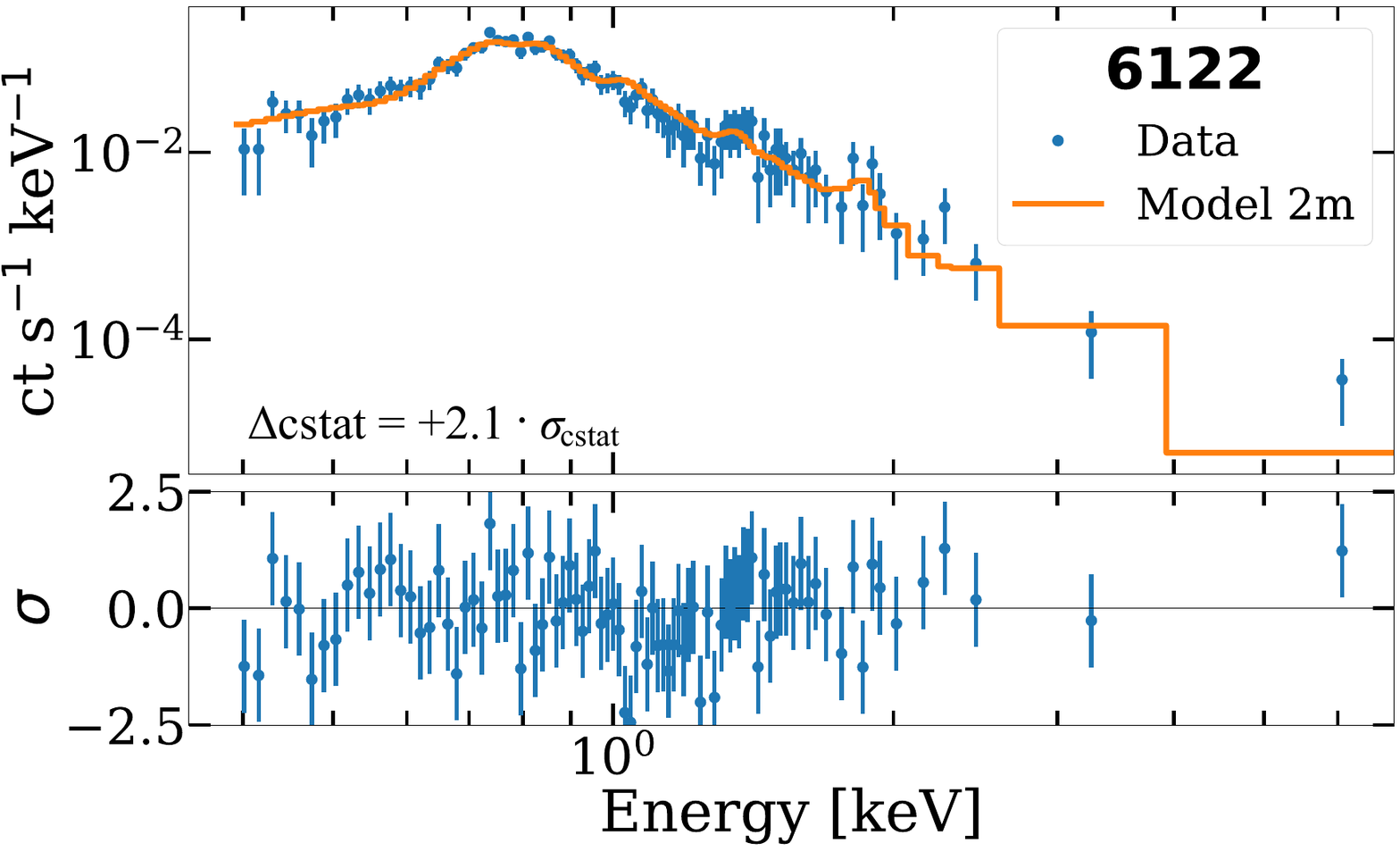} &
\includegraphics[width=0.33\textwidth,height=\textheight,keepaspectratio]{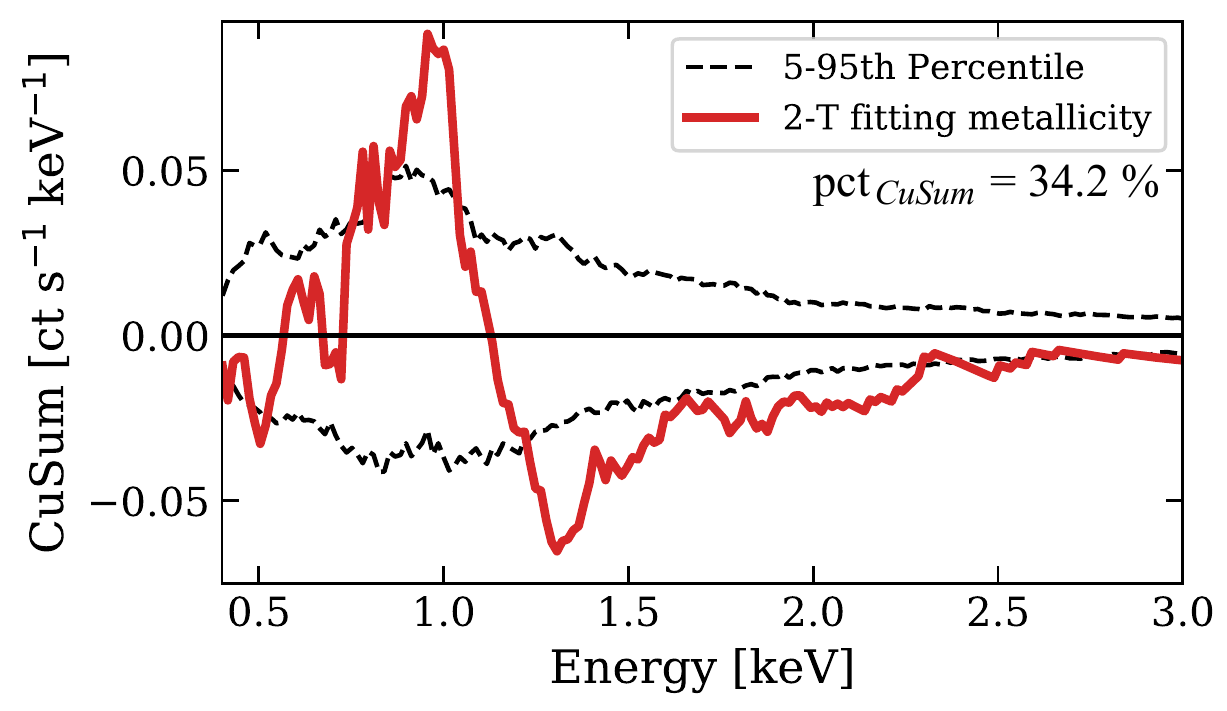} &
\includegraphics[width=0.3\textwidth,height=\textheight,keepaspectratio]{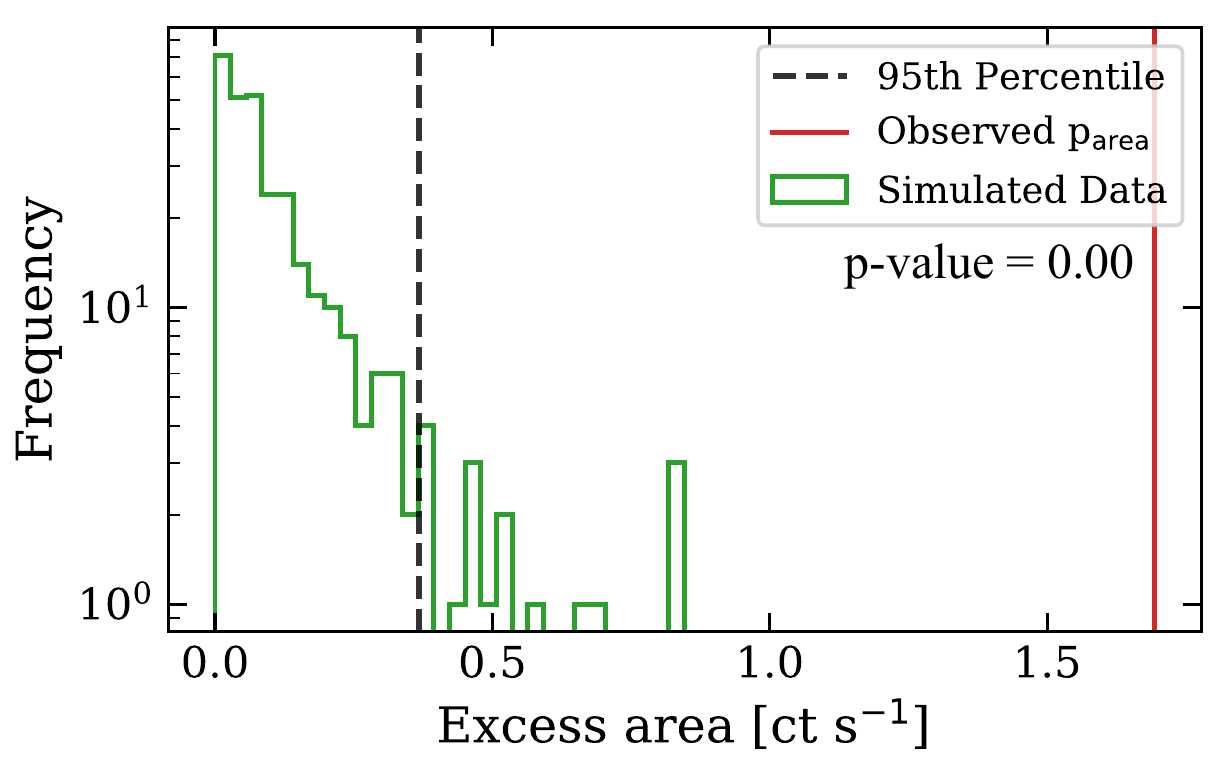} \\
\includegraphics[width=0.3\textwidth,height=\textheight,keepaspectratio]{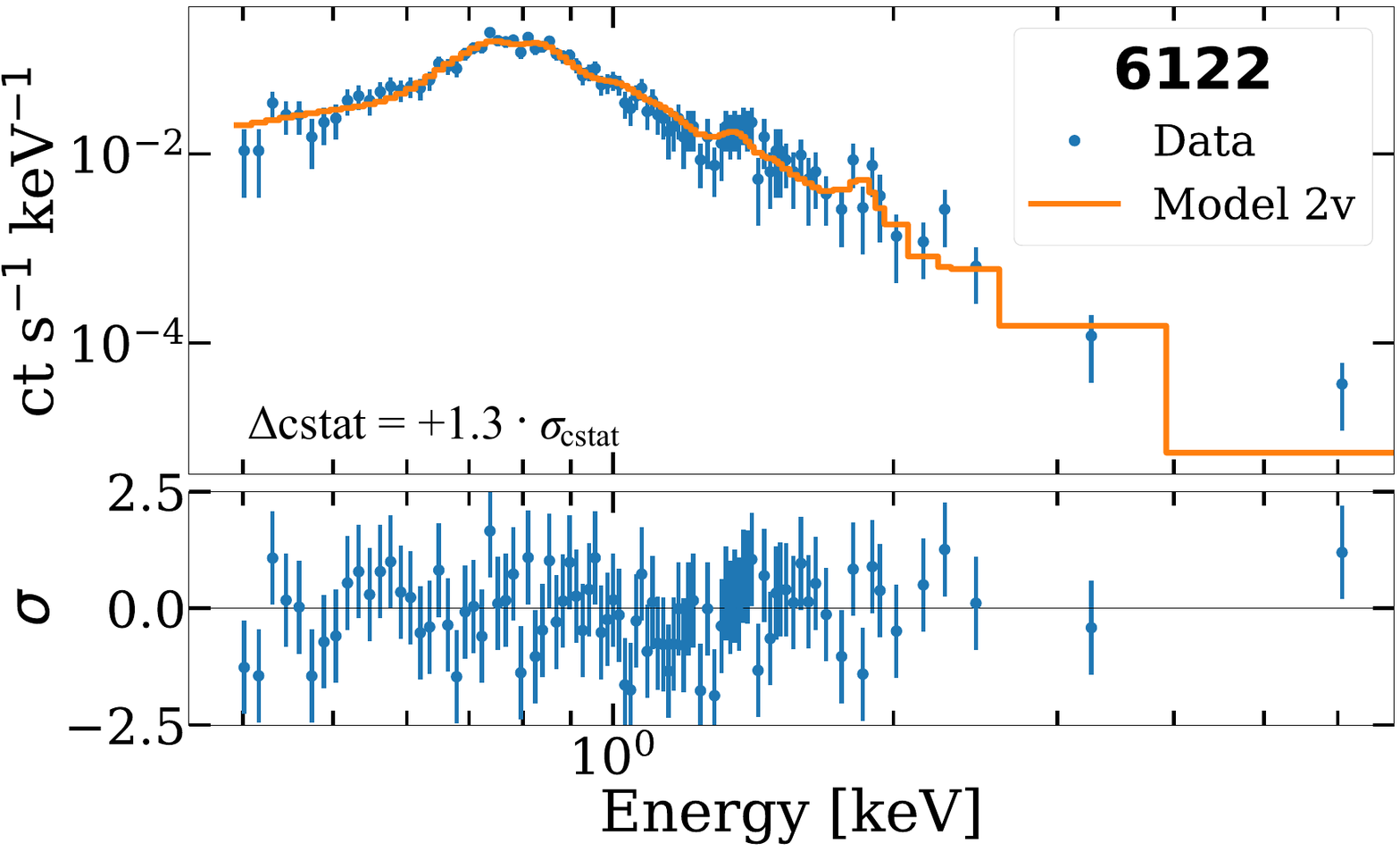} &
\includegraphics[width=0.33\textwidth,height=\textheight,keepaspectratio]{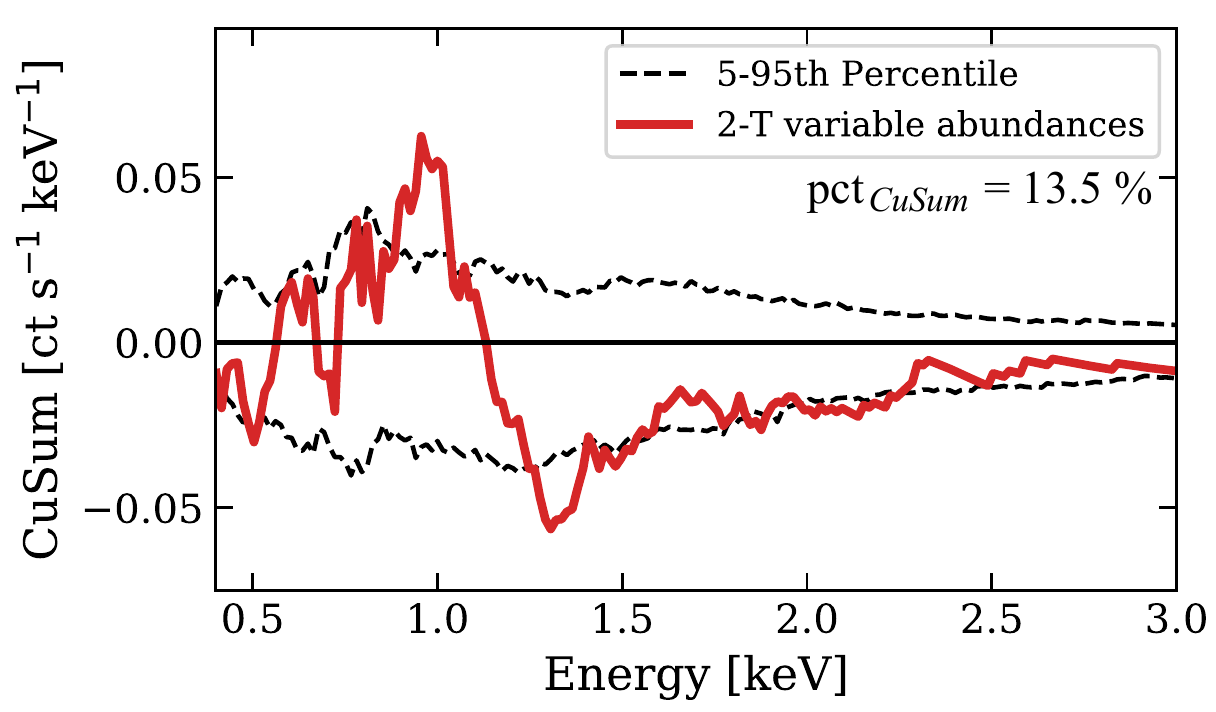} &
\includegraphics[width=0.3\textwidth,height=\textheight,keepaspectratio]{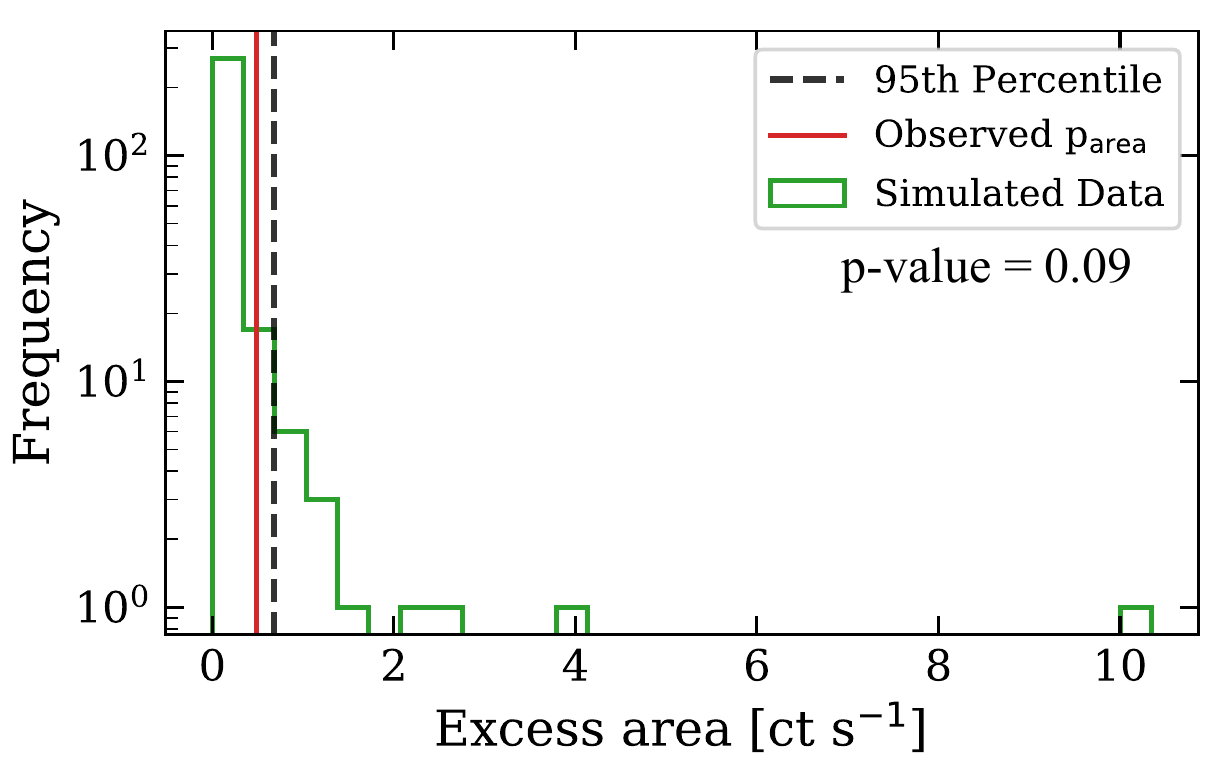} \\
\end{tabular}
\caption{Demonstrating the improvements in the measures of fit quality as model complexity increases from {\sl top} to {\sl bottom} (see text).  Each column highlights a different statistic: $\Delta{\tt cstat}$ on {\sl left}, pct$_{CuSum}$ in the {\sl middle}, and p$_\textit{area}$ on {\sl right}.  The simplest model in the top row fails in all measures; the {\tt $\Delta$cstat} value in the middle row is borderline acceptable, but the CuSum based measures are not; and in the bottom row all the measures indicate that the fit is adequate.  Increasing model complexity further risks overfitting.
}  
\label{fig:1}
\end{figure*}

In Figure~\ref{fig:1}, we show the results of the residuals analysis for three best-fit models for an exemplar dataset of HD\,179949 \citep{Acharya_2023}.  
The {\sl top} row represents a single temperature APEC model with variable metallicity (\textbf{1m}). The $\Delta\cstat = +3.3\cdot\sigma_{\tt cstat}$, indicates a poor fit. This is supported by the pct$_{CuSum} = 40.6\%$ and a p-value = 0.0.
The {\sl middle} row represents a 2-temperature model with metallicities scaling together (\textbf{2m}). The $\Delta\cstat = +2.1\cdot\sigma_{\tt cstat}$, indicates an acceptable model. The argument for improvements is solidified by pct$_{CuSum} = 34.2\%$ and p-value = 0.0.
The {\sl bottom} row represents a 2-temperature model with abundances grouped by First Ionization Potential and varying in several groups (\textbf{2v}). This gives $\Delta\cstat = +1.3\cdot\sigma_{\tt cstat}$, pct$_{CuSum} = 13.5\%$ and p-value = 0.09.  This model is therefore accepted as the best representation of the data.

pct$_{CuSum}$ is sensitive to broad differences such as would be present in a biased fit of the continuum, and p$_\textit{area}$ to small scale changes such as appear when spectral lines or abundance differences are mis-estimated.  Our method guards against overfitting to statistical fluctuations in the data by specifying criteria for acceptability that are tied to the null distributions.  We note that our arguments and method are general, and can be extended to any non-linear fitting scenario to supplement fitting procedures.

\section{Code Availability}
The code as described here is publicly available on Zenodo \citep{acharya_2023_10395978}. See \url{https://github.com/anshuman1998/csresid} for updates.

\acknowledgments
We thank Doug Burke (CXC/SDS) for suggestions with the code development. 

\bibliography{cusum}{}
\bibliographystyle{aasjournal}

\label{lastpage}
\end{document}